\UseRawInputEncoding
\documentclass[aps,preprint,groupedaddress, authoryear]{revtex4-2}
\usepackage{epsfig}
\begin{document}

%\preprint{}

\title{Synthesis of boron carbide from its elements up to 13~GPa}


\author{Amrita Chakraborti}
%\email[amrita.chakraborti@polytechnique.edu]{Your e-mail address}
%\thanks{}
\affiliation{Laboratoire des Solides Irradiés, CEA/DRF/IRAMIS, CNRS UMR 7642, École Polytechnique, Institut Polytechnique de Paris, 28 Route de Saclay, F-91120, Palaiseau, France}
\affiliation{Institut de Minéralogie, de Physique des Matériaux et de Cosmochimie (IMPMC), Sorbonne Université, UMR CNRS 7590, Muséum National d'Histoire Naturelle, IRD UMR 206, 4 Place Jussieu, 75005, Paris, France}

\author{Nicolas Guignot}
\affiliation{Synchrotron SOLEIL, Gif-sur-Yvette, France}

\author{Nathalie Vast}
\affiliation{Laboratoire des Solides Irradiés, CEA/DRF/IRAMIS, CNRS UMR 7642, École Polytechnique, Institut Polytechnique de Paris, 28 Route de Saclay, F-91120, Palaiseau, France}

\author{Yann Le Godec}
\affiliation{Institut de Minéralogie, de Physique des Matériaux et de Cosmochimie (IMPMC), Sorbonne Université, UMR CNRS 7590, Muséum National d'Histoire Naturelle, IRD UMR 206, 4 Place Jussieu, 75005, Paris, France}

\date{\today}

\begin{abstract}
The formation of boron carbide under high pressures and from elemental reactants has been studied and optimum parameters have been determined by varying the (P, T, reactants) conditions. To this end, stoichiometric mixtures of commercial $\beta$ rhombohedral boron and amorphous glassy carbon have been subjected to temperatures ranging from 1473~K to 2473~K at pressures of 2~GPa, 5~GPa and 13~GPa. Similar syntheses have been repeated for mixtures of $\beta$ boron and graphite, and amorphous boron and amorphous carbon at 2~GPa and 5~GPa. The carbon concentration of boron carbide is shown to be affected by pressure at which it is synthesised from elements, and we propose pressure as a means to control the carbon content. The formation temperature is shown to be affected by the pressure and the choice of the reactants. The effect of temperature cycling on the formation temperature has also been studied. The formation of $\alpha$~boron as an intermediate phase is seen at 5 GPa before the formation of boron carbide.

\end{abstract}

%\keywords{HPHT synthesis \sep boron carbide \sep Paris-Edinburgh press \sep multi-anvil cell}

\maketitle

%\section{}
\section{Introduction}
\label{Intro}
Boron carbide is a widely used ceramic with applications ranging from abrasives to defense purposes like safety armours and bullet-proof jackets~\cite{Caretti2008}. These applications depend on its mechanical properties such as high hardness~\cite{Herrmann2013} - its Vickers hardness is 38 GPa, a specific density as low as 2.52~g/cc~\cite{Roy2006}, and high chemical stability~\cite{Caretti2008, Kakiage2012, Thevenot1990}.

Industrially, boron carbide is formed through carbothermic or magnesiothermic reactions \cite{Roy2006, Suri2010, Alizadeh2006}, since these processes are the most viable ones economically. However, the final product obtained through these processes contains impurities. In wide contrast, producing boron carbide from elements -boron and carbon of the highest purity- results in products of the highest purity too \cite{Suri2010}. Therefore, the synthesis of boron carbide from its elements is a very important process when the purity of the final product is of primary importance.

Numerous studies have been performed on the synthesis of boron carbide with various reactants at ambient pressure or at low pressures (\textless  50 MPa) with the hot pressing technique \cite{Thevenot1990, Gosset2020}. Hot pressing results in the formation of dense boron carbide powders by mixing fine and pure powders of $\beta$ rhombohedral boron and graphite under a pressure smaller than 50 MPa and high temperatures. There have been several studies on the properties of boron carbide crystals at high pressures \cite{Yan2009, Fujii2010, Mukhanov2012, Dera2014, Hushur2016}. However, only one study has reported the synthesis of crystalline boron carbide without disorder at 8 GPa, but the synthesis reactants were not mentioned \cite{Mondal2016}. Boron carbide single crystal nanoparticles have also been synthesised in vacuum~\cite{Chen2004} and single crystalline boron carbide nanobelts have been synthesised using chemical vapour deposition (CVD) at atmospheric pressure \cite{Li-Hong2008}.

So far, only one study has recorded the synthesis parameters of boron carbide from boron and carbon at the GigaPascal (GPa) range of pressure \cite{Chakraborti2020}. However, this study was performed with limited (P, T, reactants) conditions. Also, an intermediate phase was mentioned in this paper but not identified. In particular, the reactants were either $\beta$ rhombohedral boron and amorphous carbon or $\beta$ rhombohedral boron and graphite. One of the main purposes of the current work is to fill the gap. Consequently, a systematic study of the formation of boron carbide at high pressure has been performed up to 13 GPa in order to fix the optimum parameters for synthesis. In the present work, the first objective is to identify the intermediate phase(s) that precede the formation of boron carbide and thus compete with the formation of boron carbide - this is observed at 5 GPa, but not at 2 GPa. The second objective is to find out the optimal conditions of pressure, temperature and reactants to form boron carbide. The direct synthesis from the elements avoids the formation of ternary products, which simplifies the characterisation process \cite{Anselmi-Tamburini2005}.

In the present work, the optimum (P, T, reactants) parameters of the synthesis of boron carbide from elements (boron and carbon) have been analysed for 2 GPa and 5 GPa for three combination of reactants: a) $\beta$ rhombohedral boron and amorphous carbon, b) $\beta$ rhombohedral boron and graphite and c)~amorphous boron and amorphous carbon using the Paris-Edinburgh press. Moreover, the synthesis temperature at 13 GPa has been studied using, as reactant mixture, $\beta$ rhombohedral boron and amorphous carbon using a different press with respect to Ref.~\cite{Chakraborti2020}: a multi-anvil cell. The syntheses have also been studied \textit{in situ} at the SOLEIL Synchrotron for the first time. 

In Sec.~\ref{Mat.}, the materials and methods used in the current work are presented. We report our results in Sec.~\ref{results}, and this is followed by discussions in the next section. Conclusions are drawn in Sec.~\ref{conc}.

\section{Materials and methods}
\label{Mat.}

\subsection{Synthesis conditions at 2 and 5 GPa}

The methodology of the synthesis experiments done in the Paris-Edinburgh press at 2 and 5 GPa has been discussed in an earlier work \cite{Chakraborti2020}. Additional experiments have been performed in the current work at 2 GPa and 5 GPa using the same methodology starting with a reactant mixture of amorphous boron (Pavezyum, purity \textgreater 98.5 \%, particle size \textless 250nm) and glassy amorphous carbon (Sigma-Aldrich, purity 99.95 \% trace metals basis, particle size 2 – 20 microns). 

Some \textit{in situ} experiments were also performed in the PSICHE beamline of the SOLEIL synchrotron. The \textit{in situ} experiments not only help us to confirm the results obtained from  \textit{in situ} syntheses but are also  critical for understanding the mechanism of the synthesis, especially when the intermediate phase forms. This is the first time \textit{in situ} synthesis of boron carbide from elements have been done, since the low atomic number of boron and carbon and hence the lower scattering by these elements poses a challenge. 

For these \textit{in situ} experiments, the samples were subjected to the desired pressure and then the temperature was slowly increased. At each step of the temperature increase, a diffractrogram of the sample centre was produced using the white synchrotron beam in order to understand the evolution of the reactants into the final product in real time. The white beam has an energy range of 15 - 80 keV. and it is focused to a size of 25 $\mu$m (in the vertical direction) on the sample, collimated to 50 $\mu$m in the horizontal direction. The diffractograms were taken at various sample positions in order to ensure the homogeneity of the reactions.

Energy dispersive X-ray diffraction (EDXRD) is the usual configuration used for large volume press (LVP) experiments in the PSICHE beamline. The EDXRD is complemented by the CAESAR system. The PSICHE beamline in SOLEIL synchrotron is the only synchrotron in Europe to provide the CAESAR facility, originally proposed by Wang \textit{et al.}  \cite{Wang2004}. Whereas EDXRD is advantageous in large volume press experiments in synchrotron environments because of its rapid acquisition time, background and waste signal removing efficiency, and high efficiency in working with limited sample volume and aperture, it has numerous disadvantages as well. The principal among these disadvantages are the complex corrections for intensity required for the Rietveld refinement and the presence of detector artifacts. 1D \textit{in situ}  ADXRD techniques can be used to overcome these shortcomings, but they require a special Soller slits system to remove the signal from the sample environment [88]. Thus, the combination of the complementary ADXRD and EDXRD in one system, as done in CAESAR, overcomes their respective drawbacks. The CAESAR system provides access to a higher Q range and peaks intensities are well resolved which makes the data collected suitable for structural refinement.

For all EDXRD as well as CAESAR measurements, rotating Ge solid-state detector is used. Three collimating slits are also present: one is positioned before the press and the other two are positioned between the press and the detector. 
 
The EDXRD diffractograms were each taken for 2 minutes each. The data obtained with the synchrotron radiation has been converted to Cu K$\alpha$ in order to facilitate the treatment and analysis of the data and to compare these  \textit{in situ} data to the  \textit{ex situ} results.

\subsection{ Synthesis conditions at 13 GPa}

Two syntheses at 13 GPa were also performed using a mixture of  crystalline $\beta$ boron (Prolabo, purity 99.9\% , particle size 1-10 microns) and glassy amorphous carbon which have been mixed in a hard mortar and pestle for 5 minutes in the stoichiometric ratio of 4:1. Both the \textit{in situ} and \textit{ex situ} experiments were performed using the DIA module of the multi-anvil press \cite{Liebermann2011}. A pressure of 13$\pm$0.5 GPa was attained on both cases over a span of 8 hours. 

\subsubsection{ Application of (P,T) conditions in the \textit{in situ} experiments at 13 GPa}

The \textit{in situ} experiment was performed in the PSICHE beamline of the SOLEIL synchrotron.

\begin{figure}[ht!]
\centering
%\hspace*{2.6cm} 
\includegraphics[width=0.5\textwidth]{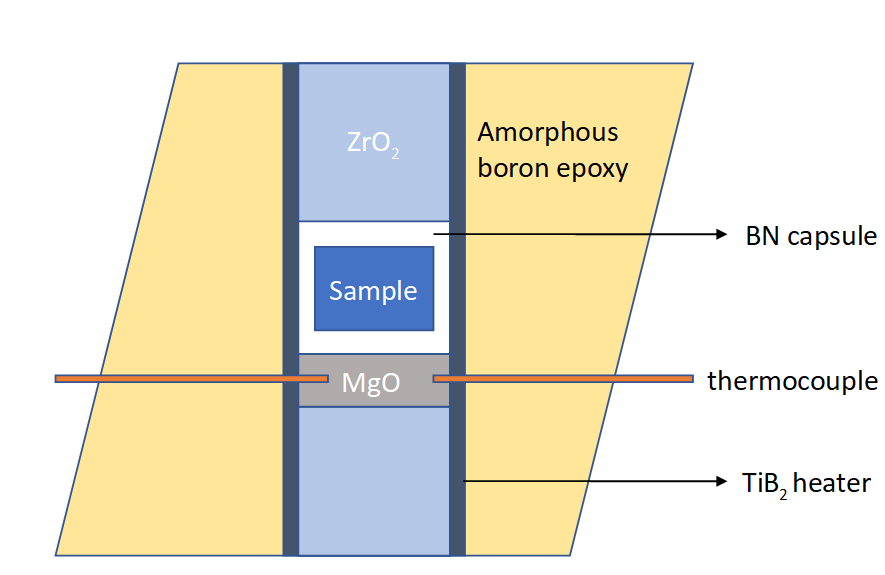}
\caption{\label{fig:MAP_assembly_LaCrO3}Set-up 1: The sample assembly used for the \textit{in situ} multi-anvil experiment. TiB$_{2}$ heater was used inside an amorphous boron epoxy octahedron. The sample capsule is made of hexagonal boron nitride (hBN).}
\end{figure}

For the \textit{in situ} experiment, the reactant mixture was put in the set-up (set-up 1) shown in figure~\ref{fig:MAP_assembly_LaCrO3}. Pressure was estimated using MgO as a pressure calibrant \cite{Dewaele2000, Tange2009}.

The temperature for the \textit{in situ} experiment was not increased uniformly (i.e at a constant rate) to the final dwell temperature, as was done in the case of the other experiments. Instead, a thermal cycling was applied in order to observe its effects. The temperature was increased to a specific value for around 3 minutes, the EDXRD pattern was recorded in the center of the sample, and in the following step, the temperature was dropped to 300 K. Next, the temperature was raised to the next higher value and the same process was repeated. The upper limit of the temperature that was attained in each successive cycle has been recorded in table~\ref{tab:temcycvalues}. After the final quenching to 300 K, the sample was then decompressed slowly over 4 hours. Finally, a diffractogram was taken using the CAESAR system.

\begin{table}[htp]
\begin{center}
\begin{tabular}{|ccc|}
\hline\hline
Cycle  & Lower Temp. & Higher Temp. \\
\hline\hline
%1 & 300 K & 508 K \\
1 & 300 K & 875 K \\
2 & 300 K & 1170 K \\
3 & 300 K & 1360 K \\
4 & 300 K & 1670 K \\
5 & 300 K & 1912 K \\
6 & 300 K & 2430 K \\
\hline
\end{tabular}
\end{center}
\caption{The upper and lower values of temperature attained during each cycle in the \textit{in situ} run.}
\label{tab:temcycvalues}
\end{table}

\subsubsection{ Application of (P,T) conditions in the \textit{ex situ} experiments at 13 GPa}

\begin{figure}[ht!]
%\centering
%\hspace*{2.6cm} 
\includegraphics[width=0.5\textwidth]{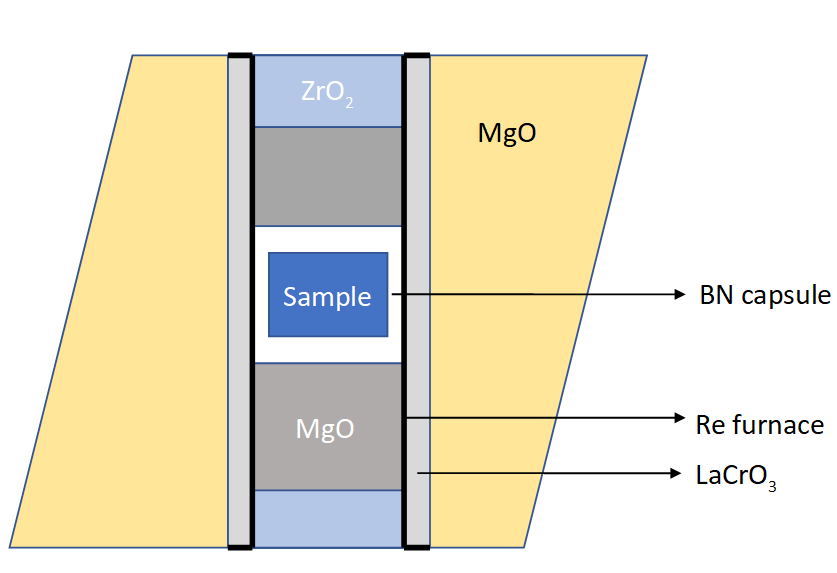}
\caption{\label{fig:MAP_assembly_Re}Set-up 2: The sample assembly used for the \textit{ex situ} multi-anvil experiment. Rhenium foil was used as a heater instead of the LaCrO$_{3}$ used in the \textit{in situ} run.}
\end{figure}

For the \textit{ex situ} experiment, a different type of set up was used (set-up 2), as shown in figure~\ref{fig:MAP_assembly_Re}. The main differences are the use of rhenium foil as the heater and LaCrO$_{3}$ sleeve as a thermal insulator. The metal heater enables us to have a larger sample volume compared to set-up 1, which is important for subsequent \textit{ex situ} characterisation. 
The temperature was raised slowly and constantly at a rate of 2 K/s. Temperature was measured using a type~C thermocouple (W 5 \% Re/ W 26 \% Re) with an error of evaluation of $\pm$~20 K. Then, the sample was subjected to a dwell time of 30 minutes to the \textit{ex situ} sample: no thermal cycling was performed on this sample. 

Pressure was fixed at 13~GPa by using a calibration curve determined beforehand, with an error of evaluation of $\pm$ 0.2 GPa.

After the dwell time of 30 minutes, the sample was quenched rapidly by removing the electrical power. The decompression was done slowly over the span of 4 hours. The sample was then recovered carefully and subjected to characterisation by \textit{ex situ} X-ray diffraction.

\section{Results}
\label{results}

\subsection{\textit{Ex situ} and \textit{in situ} diffraction at 2 and 5 GPa}

We define the synthesis temperature as the lowest temperature where we observe the formation of boron carbide. The optimal temperature, on the other hand, is defined as the lowest observed temperature at which boron carbide is the prominent product.

Further syntheses performed in this work has confirmed the preliminary results of our previous work \cite{Chakraborti2020} to find out the synthesis temperatures and optimal temperatures of boron carbide at 2 and 5 GPa using reactant mixtures of a) $\beta$ boron and amorphous carbon, and b) $\beta$ boron and graphite. Moreover, in the current work, we have further expanded the scope of the study by using other reactants (amorphous boron), higher pressure values, temperature cycling as well as \textit{in situ} syntheses. The results have been tabulated in table~\ref{tab:synthesistemperatures}.

\begin{figure*}[ht!]
%\centering 
\hspace*{-1.5cm}
\includegraphics[width=0.8\textwidth]{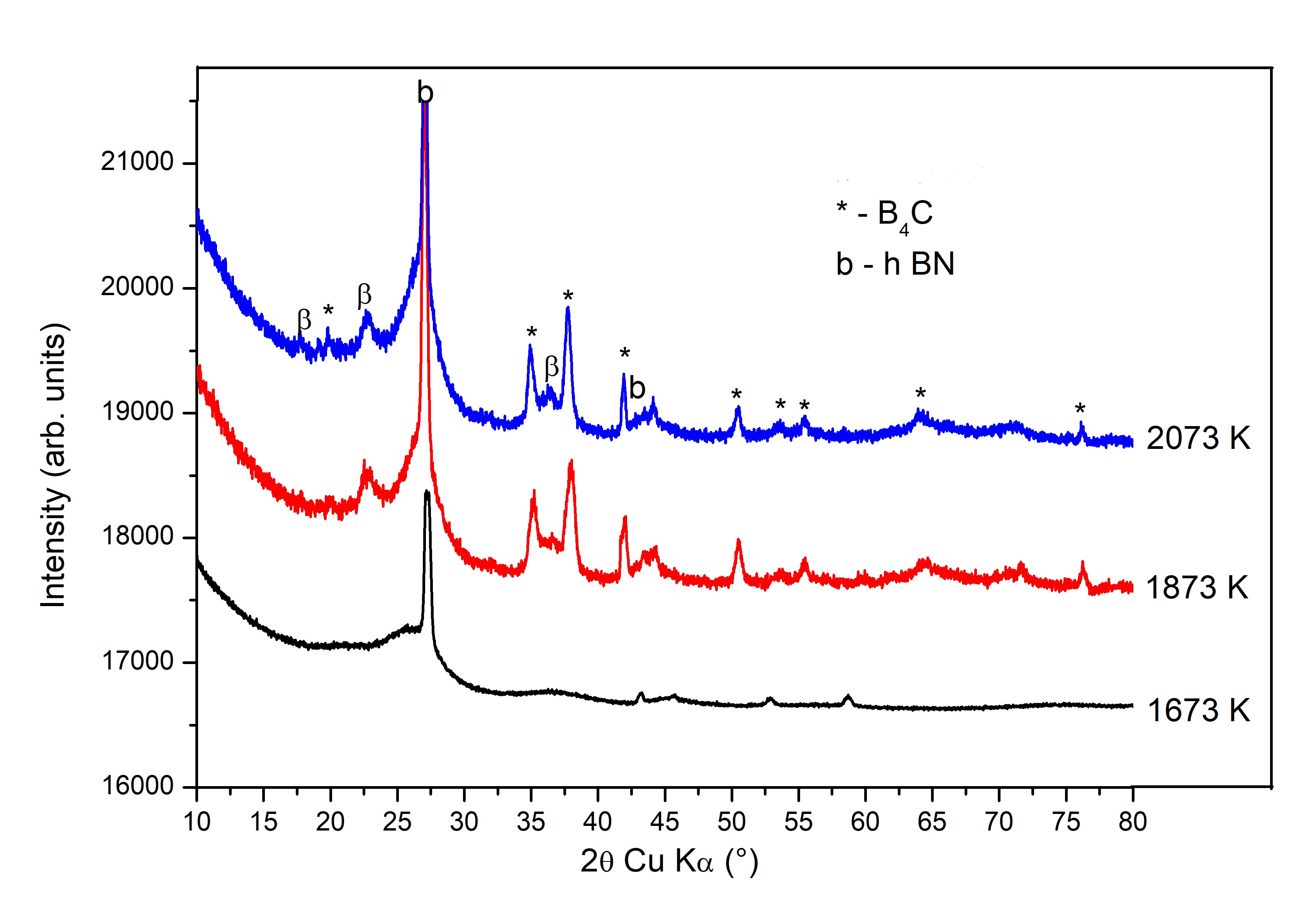}
\caption{\label{fig:2GPa_aB+C} \textit{Ex situ} X-ray diffraction pattern of quenched sample from the mixture of amorphous boron  and amorphous carbon under 2 GPa for 2 hours.  It is shown that boron carbide \cite{boroncarbide} is formed at 1873 K with the amorphous boron reactant. No $\alpha$~boron is formed. The hBN peaks come from the capsule enclosing the sample, and not the sample itself.}
\end{figure*}

In the current work, the mixture of amorphous boron and amorphous carbon have been subjected to a pressure of 2 $\pm$ 0.2~GPa and 5 $\pm$ 0.2~GPa respectively. At 2 GPa, the synthesis temperature of boron carbide is 1873 $\pm$ 20 K. In figure~\ref{fig:2GPa_aB+C}, we report the XRD pattern showing the synthesis of boron carbide from amorphous boron and amorphous carbon at 2 GPa.

\begin{figure*}[ht!]
%\begin{figure}[H]
%\centering 
\hspace*{-1.5cm}
\includegraphics[width=0.8\textwidth]{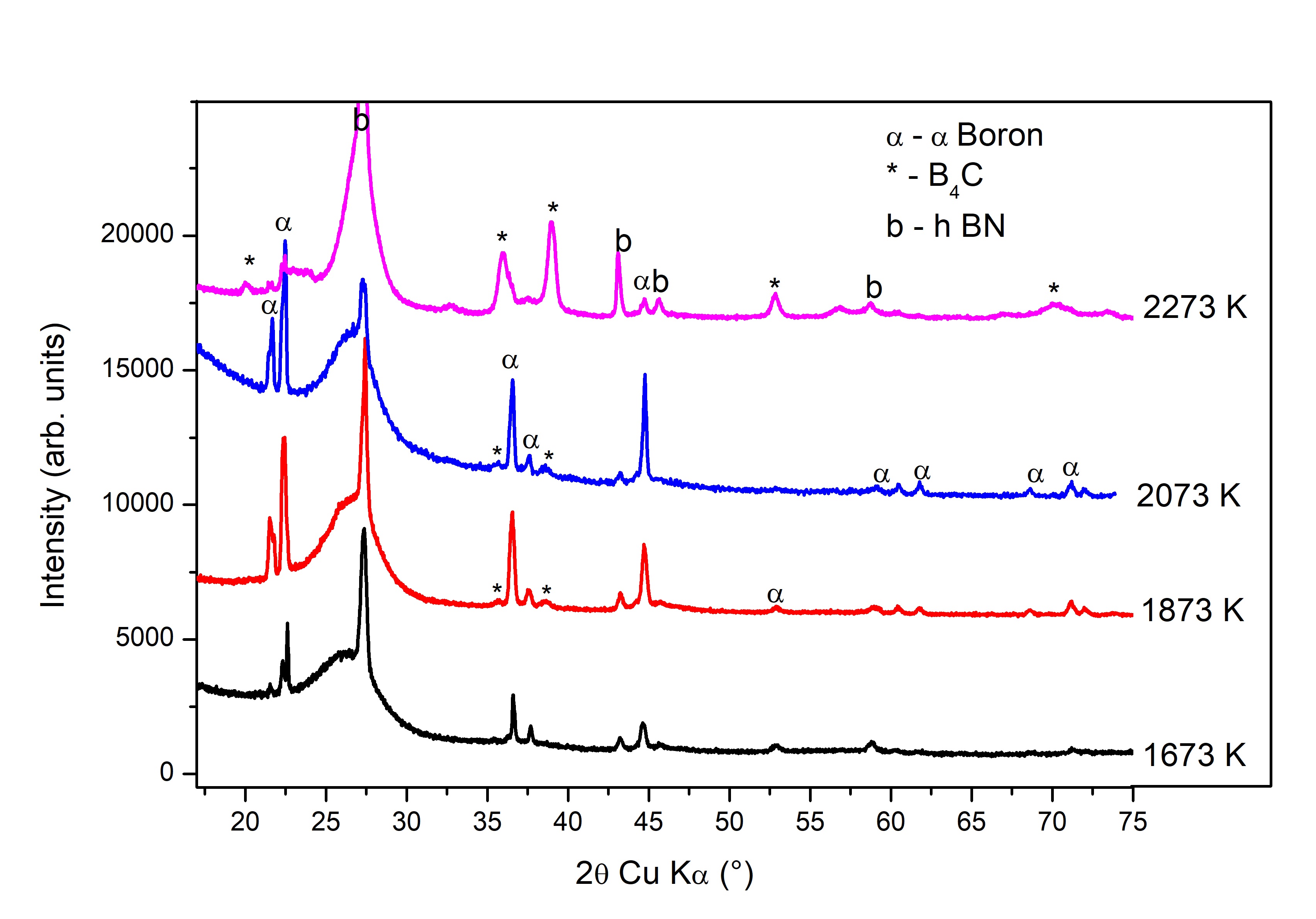}
\caption{\label{fig:5GPa_aB+C} \textit{Ex situ} X-ray diffraction pattern of quenched sample from the mixture of amorphous boron  and amorphous carbon under 5 GPa for 2 hours.  It is shown that boron carbide \cite{boroncarbide} starts forming at 1873 K with the amorphous boron reactant, with $\alpha$~boron remaining the prominent phase. Boron carbide becomes the prominent phase at 2273 K. The hBN peaks come from the capsule enclosing the sample, and not the sample itself.}
\end{figure*}

At 5 GPa as well, the temperature of formation of boron carbide decreases from 2273 $\pm$ 20 K (when $\beta$~boron is used as a reactant) to 1873 $\pm$ 20 K.  In figure~\ref{fig:5GPa_aB+C}, the XRD patterns at various temperatures for the 5 GPa series are shown. It is observed that only very small XRD peaks of boron carbide appear at temperatures of synthesis of 1873 K and 2073 K. At these temperatures (1873 K and 2073 K), an intermediate phase, identified as $\alpha$~boron remained the most prominent phase. Boron carbide becomes the most prominent phase at 2273 K, similar to what was observed during the syntheses with $\beta$~boron and amorphous carbon \cite{Chakraborti2020}. Moreover, similar to the case of synthesis from $\beta$~boron, no formation of $\alpha$~boron was recorded for the syntheses with amorphous boron at 2 GPa. This is in line with what was observed for the syntheses with $\beta$~boron and amorphous carbon in Ref.~\cite{Chakraborti2020}.

\begin{figure*}[ht!]
%\begin{figure}[H]
%\centering 
\hspace*{-1.5cm}
\includegraphics[width=0.8\textwidth]{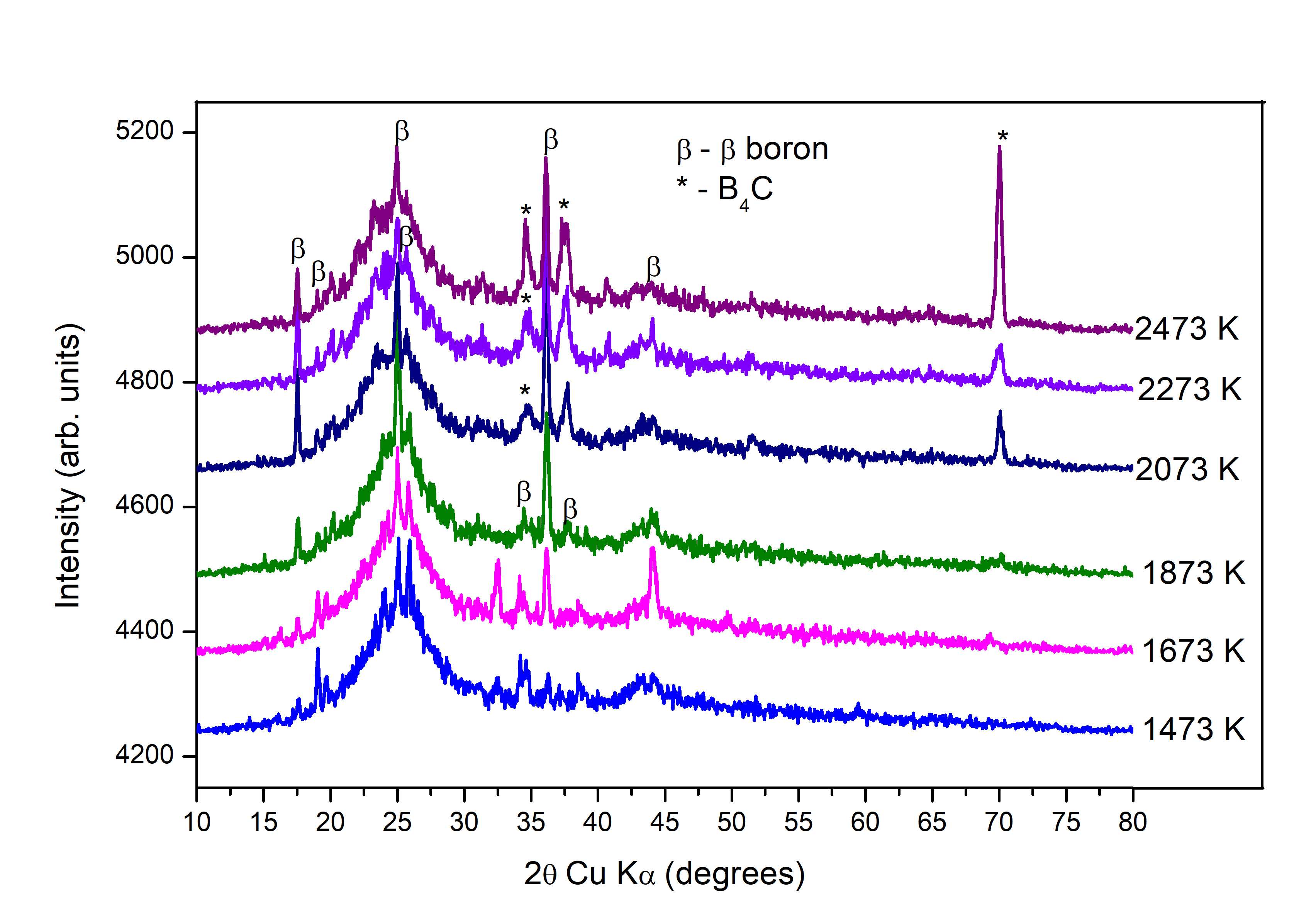}
\caption{\label{fig:2GPa_B+C_Soleil} \textit{In situ} diffraction pattern taken at the center of the sample of a mixture of crystalline β boron and amorphous carbon at 2 GPa. The data obtained from the synchrotron radiation was converted to that of Cu K$\alpha$ radiation.}
\end{figure*}

Additionally, \textit{in situ} synthesis at 2 GPa, in figure~\ref{fig:2GPa_B+C_Soleil}, with a reactant mixture of $\beta$ boron and amorphous carbon, corroborated the  \textit{ex situ} results reported in \cite{Chakraborti2020} - boron carbide forms at 2073 K from a mixture of $\beta$ boron and amorphous carbon under a pressure of 2 GPa. The diffractograms were taken within 5 minutes of raising the temperature each time, unlike the 2 hour dwell time that the samples were subjected to for the \textit{ex situ} syntheses. This means that the reactions were still incomplete when the diffractograms were taken - this is the reason behind the presence of $\beta$ boron and amorphous carbon in the diffractograms even at high temperatures.

\begin{table*}[h!]
%\begin{table*}[H]
%\begin{center}
%\hspace*{-2.0cm}
\begin{tabular}{|c|c|c|c|c|c|}
\hline\hline
Reference                            &  Pressure          &  Reactants  & heating& Synthesis & Optimal      \\
&(GPa)&&rate&temp. (K)&temp. (K) \\
\hline\hline
Wang  \textit{et al}, 2009 \cite{Wang2009}          &  0.02                     &    B + graphite &   constant          &     1573 &    -                       \\\hline
Roszeitis  \textit{et al}, 2014 \cite{Roszeitis2014} &  0.05                    &    B + graphite & constant           &     1606  &   -              \\ \hline
This work  &	2 $\pm$ 0.2 	&	amorph. B + amorph. C	& constant	&		1873 $\pm$ 20 	& 1873 	\\ \hline
This work  &	5 $\pm$ 0.2 	&	amorph. B + amorph. C	& constant	&		1873 $\pm$ 20 	& 2273 	\\ \hline
Our work, 2020 \cite{Chakraborti2020}   &     2 $\pm$ 0.2 	         &    $\beta$~B + amorph. C&	constant              &     2073 $\pm$ 20 &	 2073         \\\hline
Our work, 2020 \cite{Chakraborti2020}    &    2 $\pm$ 0.2 		 &   $\beta$~B + graphite&	constant	  &        2273 $\pm$ 20 	& \textgreater 2273		       \\ \hline
Our work, 2020 \cite{Chakraborti2020}    &    5 $\pm$ 0.2 		 &   $\beta$~B + amorph. C&	constant	  &        2273 $\pm$ 20 & 2273	   \\\hline
Our work, 2020 \cite{Chakraborti2020}   &	5 $\pm$ 0.2 		&	 $\beta$~B + graphite&	constant	&		2473 $\pm$ 20 	& \textgreater 2573		\\\hline
This work & 13 $\pm$ 0.5  & $\beta$~B + amorph. C & cycling & 2430 $\pm$ 20 & \textgreater 2430 \\\hline
%&&&(temperature cycling)&\\\hline
This work & 13 $\pm$ 0.2 &  $\beta$~B + amorph. C & constant & \textgreater 2473 & \textgreater 2473 \\\hline
%&&&(constant heating)&\\\hline
\end{tabular}
%\captionsetup{width=17cm}
\caption{\label{tab:synthesistemperatures} Summary of the synthesis temperatures and optimal temperatures of formation of boron carbide from elemental boron and carbon, depending on the pressure and on the solid-state form of carbon and boron, in ascending order of synthesis temperature and pressure.}
%\end{center}
\end{table*}

\subsection{Diffraction at 13 GPa}

We also report the results at 13 GPa.

\begin{figure*}[ht!]
%\begin{figure}[H]
%\centering 
%\hspace*{-1.6cm}
\includegraphics[width=0.8\textwidth]{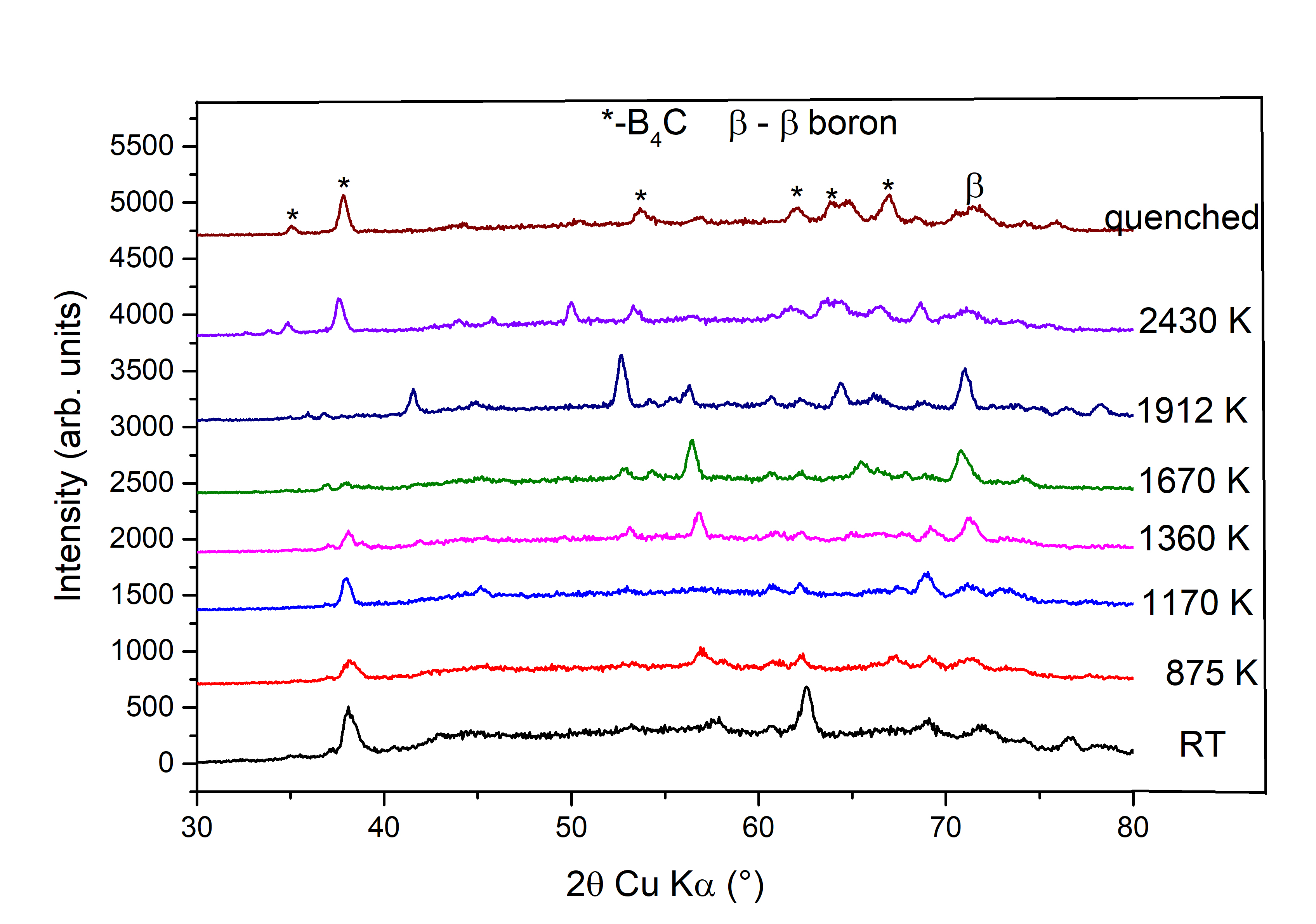}
\caption{\label{fig:EDXRD_MA77}  Temperature cycling at 13 GPa: \textit{In situ} diffraction pattern taken at the center of the sample of a mixture of crystalline $\beta$ boron and amorphous carbon at 13 GPa for various temperatures during the \textit{in situ} run. The data obtained from the synchrotron radiation was converted to that of Cu K$\alpha$ radiation. All the unmarked peaks are those of crystalline $\beta$ boron.}
\end{figure*}

During the \textit{in situ} run at 13 GPa, a series of EDXRD patterns were taken immediately after the temperature was raised in each cycle and before the system was quenched. Figure~\ref{fig:EDXRD_MA77} shows the EDXRD patterns as a function of rising temperature at 13 GPa. In the figure, the data obtained with the synchrotron radiation has been converted to Cu K$\alpha$ in order to facilitate the treatment and analysis of the data. 

As the upper limit of temperature is increased in each cycle, only some $\beta$ boron peaks are observed up to 1912 K. Some peaks of $\beta$ boron are different from one cycle to another: this can be attributed to the recrystallisation of $\beta$ boron under high pressure and temperature conditions \cite{Parakhonskiy2011}.

However, in the last cycle, as temperature is increased to 2430 K, most of the $\beta$ boron has reacted and the formation of boron carbide is observed. 

After the last cycle of temperature increase, the system is quenched and EDXRD is performed again to observe whether boron carbide remains in the system. As the final EDXRD pattern in figure~\ref{fig:EDXRD_MA77} shows, the boron carbide peaks have remained unchanged after quenching.

\begin{figure*}[ht!]
%\begin{figure}[H]
%\centering 
%\hspace*{-1.5cm}
\includegraphics[width=0.8\textwidth]{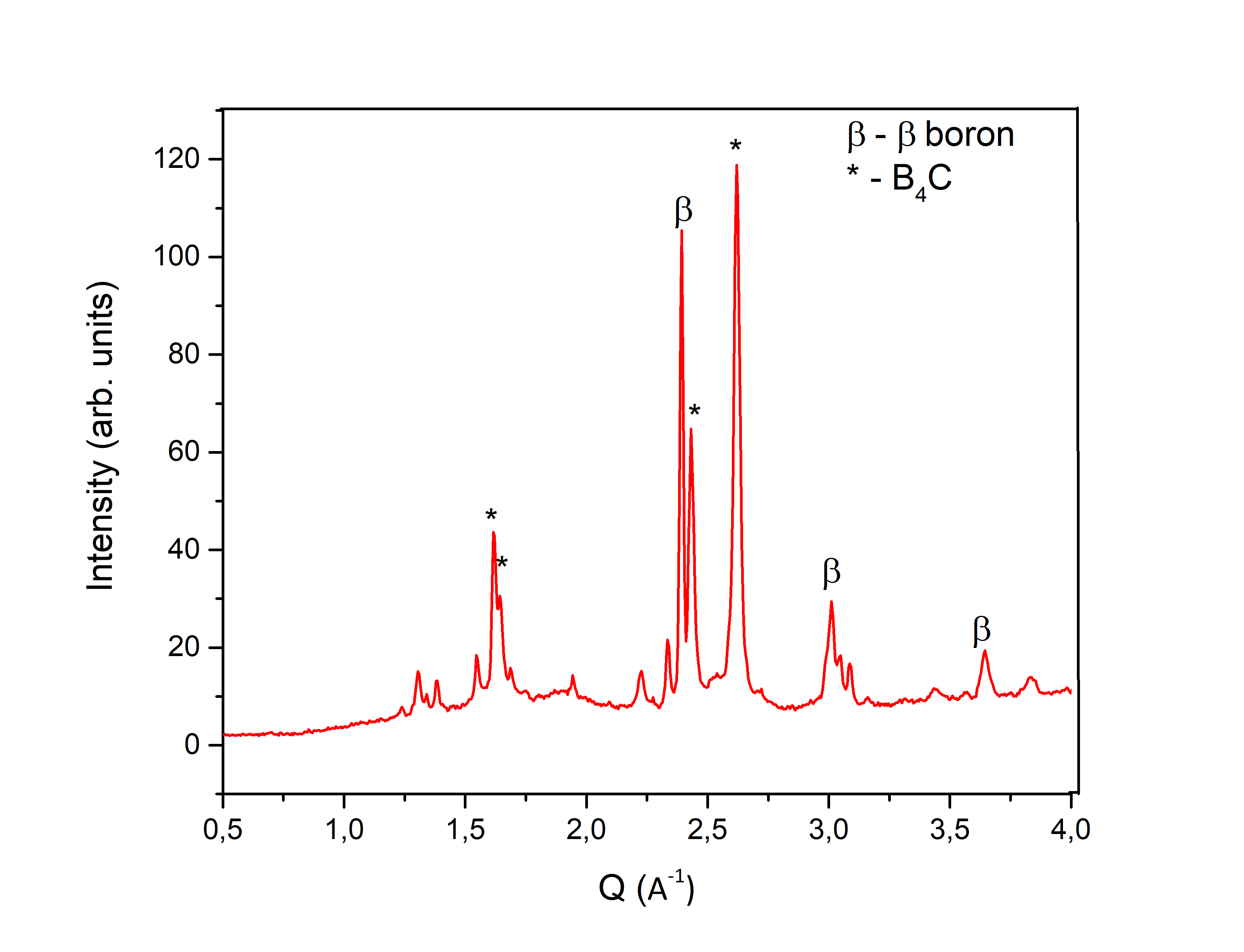}
\caption{\label{fig:Caesar_MA77} Temperature cycling at 13 GPa: \textit{In situ} diffraction pattern taken at the center of the sample of a mixture of crystalline $\beta$ boron and amorphous carbon at ambient conditions after applying a pressure of 13 GPa and temperature cycling during the \textit{in situ} run. The data was obtained from the synchrotron radiation using the CAESAR system~\cite{Yang2008}. All the unmarked peaks are those of crystalline $\beta$ boron. }
\end{figure*}

After this, the sample was decompressed and another XRD pattern was taken using the CAESAR system. The CAESAR pattern is shown in figure~\ref{fig:Caesar_MA77}: it is confirmed that boron carbide has remained in the sample after the process.

%\newpage

\begin{figure*}[ht!]
%\begin{figure}[H]
%\centering 
\hspace*{-1.5cm}
\includegraphics[width=0.8\textwidth]{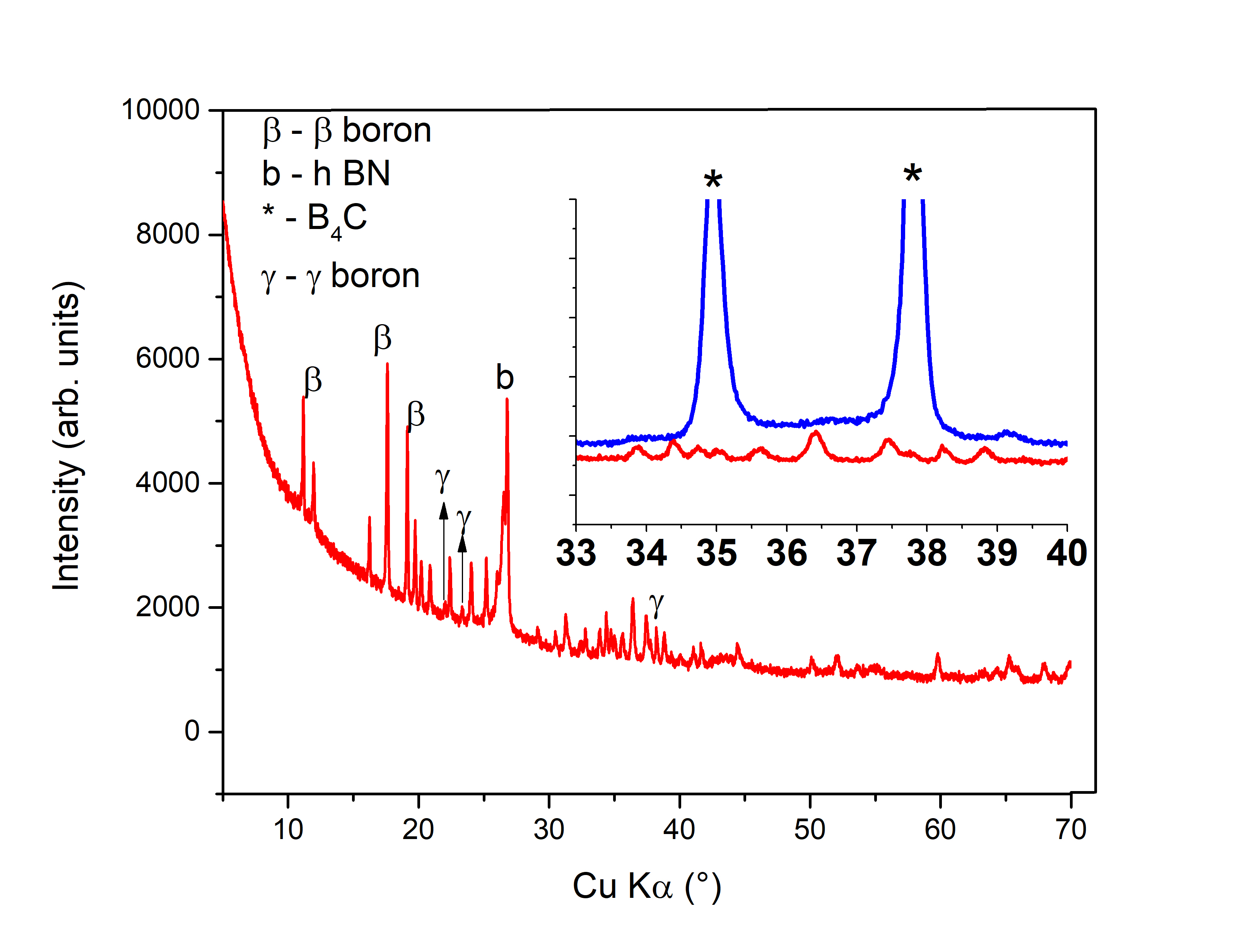}
\caption{\label{fig:MA_exsitu} Constant heating rate at 13 GPa: X-ray diffraction pattern of quenched sample from the mixture of crystalline $\beta$ boron \cite{betaboron} and amorphous carbon under 13~GPa at 2473~K for 1 hour. The inset is the zoom between 33 ° and 40 ° of the superposition of the curve of the product (same colour as main plot) and the diffractogram of conventional B$_{4}$C. %taken from figure~\ref{fig:2GPa_B+C}. 
It is shown that no new peaks corresponding to boron carbide \cite{boroncarbide} form in the sample. The hBN peaks come from the capsule enclosing the sample, and not the sample itself.}
\end{figure*}

In order to verify whether the temperature cycling had any effect on the formation of boron carbide, a second experiment was performed. In this \textit{ex situ} experiment, the temperature was raised to 2473 K slowly without any cycling. The XRD performed on the recovered sample after the experiment is shown in figure~\ref{fig:MA_exsitu}: no boron carbide peaks have been observed in the sample. In order to establish this observation, the XRD diffractogram has been compared to that of boron carbide formed at 2473 K under 2 GPa in the inset of figure~\ref{fig:MA_exsitu}. The 2$\theta$ angle domain between 33° - 40° (in inset of figure~\ref{fig:MA_exsitu})  shows the most prominent peaks of boron carbide (012) and (104) under Cu K$\alpha$ radiation. As is evident from the comparison, the sample quenched after a HPHT treatment at 13~GPa and 2473~K does not contain any boron carbide. Some trace amounts of $\gamma$ orthorhombic boron~\cite{Oganov2009} was found in the sample, but no $\alpha$ boron was formed, contrary to the case at 5 GPa. At 13 GPa, after heating to 2473 K, the quenched sample is overwhelmingly composed of $\beta$ rhombohedral boron.

\subsection{Raman spectroscopy at 13 GPa}

\begin{figure*}
%\begin{figure}[H]
%\centering 
\hspace*{-1.5cm}
\includegraphics[width=0.8\textwidth]{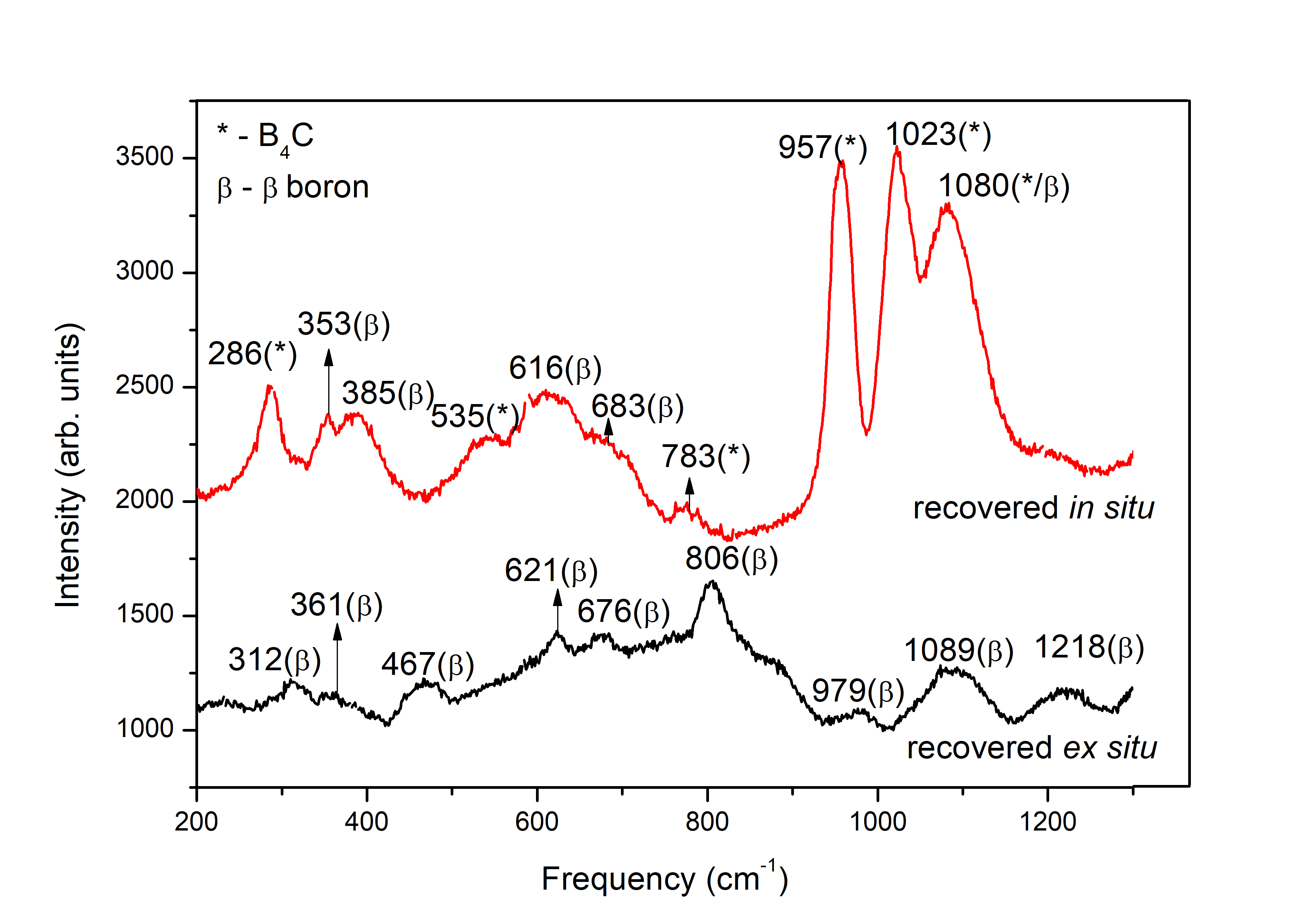}
\caption{\label{fig:multi_anvil_Raman} Raman spectroscopy of recovered samples from the \textit{in situ} (temperature cycling) and  \textit{ex situ} (constant heating) synthesis at 13~GPa. The \textit{ex situ} sample shows only $\beta$ boron peaks after HPHT treatment, while the  \textit{in situ} sample shows both $\beta$ boron and boron carbide peaks. }
\end{figure*}

Figure~\ref{fig:multi_anvil_Raman} shows the results of the Raman spectroscopy done on the recovered samples from the \textit{in situ} and  \textit{ex situ} synthesis. Crystalline $\beta$ boron peaks were found in both samples, while the  \textit{in situ} synthesis had additional boron carbide peaks. No peaks corresponding to $\gamma$ boron or $\alpha$ boron were found in either of the samples.

%\begin{figure}[ht!]

% \begin{figure}
% \end{figure}

\section{Discussions}
\label{disc}

\subsection{Search for the optimal synthesis parameters}
The synthesis at 2 GPa follows a well-understood and direct path, similar to what has been reported in the literature \cite{Wang2009, Roszeitis2014}  for the low pressure syntheses. However, the synthesis temperature rises considerably. A temperature increase by more than 250 K with respect to that of ambient pressure is required to form boron carbide at 2 GPa. Only by 1873 K (in the case of amorphous boron and amorphous carbon), 2073 K (in the case of $\beta$~boron and amorphous carbon) and 2273 K (in the case of $\beta$~boron and graphite) do we find that boron carbide has formed. 

Regarding the optimisation of the synthesis conditions, amorphous carbon turns out to be a better reactant than graphite for synthesising boron carbide at elevated pressures, since the use of graphite requires a further increase of the synthesis temperature. This is attributed to the higher reactivity of amorphous carbon because of its high concentration of dangling bonds \cite{Kouchi2011, Caro2018, Robertson1986}. Also, carbon atoms have higher mobility in amorphous carbon than in graphite \cite{Limandri2019}. The higher mobility of the carbon atoms increases the diffusion of carbon throughout the reacting mixture - this allows the reactants to come in contact with each other thus leading to synthesis of boron carbide at lower temperatures. 

Amorphous boron is also found to be a better reactant than $\beta$~boron - this can also be attributed to the better reactivity of amorphous boron compared to crystalline boron due to the absence of an ordered crystalline structure \cite{Portehault2016}, \cite{Yang2013}. %However, at 5 GPa, the reaction with both amorphous boron and $\beta$~boron produce $\alpha$~boron as the prominent product up to 2073~K. This indicates that $\alpha$~boron is an intermediate step towards the formation of boron carbide from amorphous or $\beta$~boron and amorphous carbon under a pressure of 5~GPa.

Moreover, the higher reactivity of amorphous boron and carbon can also be attributed to the higher surface area available in amorphous substances leads to greater contact area for reactions to take place, compared to their crystalline counterparts.

At 5 GPa, the reaction with both amorphous boron and $\beta$~boron produce $\alpha$~boron as the prominent product up to 2073~K. This indicates that $\alpha$~boron is an intermediate step towards the formation of boron carbide from amorphous or $\beta$~boron and amorphous carbon under a pressure of 5~GPa.

The synthesis temperature also increases with the increase of pressure in all of the cases except for that of amorphous boron as reactant. Yang \textit{et al.}~\cite{Yang2013} have suggested that high pressure can decrease the reaction efficiency of boron greatly. In their experiments, the evaporation rate of the surface boron oxide would decrease as the pressure increased. However this does not explain the reason behind the increase of synthesis temperature when amorphous carbon is replaced by graphite, as the same crystalline $\beta$~boron was used in both cases. Another possible explanation is that high pressure may inhibit the diffusion of carbon atoms through the interstitial spaces in crystalline boron and thus, it would slow down the reaction. The application of higher temperature values would then favour the diffusion process that leads to the formation of boron carbide. In another study, it has also been noted that an inverse relationship between pressure and the reaction constant exists at the high pressure limit for chemically activated reactions \cite{Carstensen2007}. Therefore, as the pressure applied during the synthesis increases, the reaction rate falls. This effect can be compensated by an increase in temperature, which results in the increased synthesis temperature of boron carbide. Thus, temperature can be said to act as negative pressure~\cite{Jay2019}, and so higher temperature at higher pressure may be the signature that the change in diffusion due to pressure is compensated by the change in diffusion due to temperature. This relationship between the rate of change of the diffusion constant with temperature and pressure in the case of formation of boron carbide from its elements under pressure has been expressed in equation~\ref{Eq: diffusion_constant}.

\begin{equation}
\label{Eq: diffusion_constant}
%\Bigg[ - \frac{\nabla^{2}}{2} + V_{n}(r) + V_{H}(r)+ V_{x}(r)+ V_{c}(r)\Bigg] \phi_{i}(r) = \epsilon_{i} \phi_{i}(r)
%V_{X}(r) = V_{X}(n)
\frac{\partial D}{\partial T} = - k \frac{\partial D}{\partial P}
\end{equation}

where D is the diffusion constant of carbon or boron atoms, T is the temperature, P is the pressure and k is the constant of proportionality.

\subsection{Crystallite size and particle size}
The average crystallite size of the boron carbide as synthesised at 2 and 5 GPa has been determined using the Scherrer formula \cite{Patterson1939}. The average crystallite size (about 25 nm) was found  to decrease with pressure, and to increase with the synthesis temperature. %This was explained by the restricted growth of the crystal as a result of increased pressure. The increase in temperature counteracts this effect. 
The size was also found to slightly decrease when graphite is used as a reactant instead of amorphous carbon. %This was attributed to a larger carbon mobility in amorphous carbon than that in graphite \cite{Limandri2019}. 
Concerning the granulometry of the powder, scanning electron microscopy showed a particle size of 3 - 6 microns irrespective of reactant mixture or synthesis temperature. 

\subsection{Carbon content of the synthesised boron carbide}
It is also worthwhile to try to find out the carbon concentration of the boron carbide synthesised using different reactants and different pressures. It is difficult to do so using XRD analysis techniques, because of the close atomic numbers of boron and carbon \cite{Gosset1991}. Chemical analysis has been used extensively in literature to find out the carbon concentration in boron carbides  \cite{Schwetz1991, Aselage1992}. However, chemical analysis can overestimate the carbon present in the samples due to the presence of free carbon in boron carbides. In the present work, we have deduced the approximate carbon content of the boron carbides synthesized through their lattice parameters \cite{Gosset2020, Aselage1992}. In the literature, the carbon concentration was deduced as a function of three parameters - the hexagonal cell parameters $a$ and $c$ and the rhombohedral cell volume $V$. The lattice parameters for the syntheses were determined by the Le Bail refinement using the software Powdercell \cite{Kraus1996}. The unit cell volume values were then compared with the reference values in the literature \cite{Gosset2020, Aselage1992} and an estimate of the carbon content was deduced (table~\ref{tab:carbonconcentration}). These analyses were done for the boron carbides synthesised from amorphous carbon at the temperatures where boron carbide first appears. 

Results show that the carbon content in boron carbides decreases with increase of the pressure of synthesis. This can be attributed to the lower diffusion of carbon in boron as the pressure increases, leading to the incorporation of a smaller amount of carbon in the unit cells: this argument also holds true for the rise of synthesis temperature with increase in pressure, for the same reasons as given above i.e temperature acts as negative pressure.

Thus, we propose pressure as a way to modify the carbon concentration of boron carbides.

%\begin{widetext}
\begin{table*}[ht!]
%\begin{table*}[H]
%\begin{landscape}
%\begin{center}
%\hspace*{-1.5cm} 
\begin{tabular}{|c|c|c|c|c|c|c|}
\hline\hline
P (GPa)                           &  T (K)        &  Reactants  &  $a$ (\AA) & $c$ (\AA) & $V$ (\AA$^3$) & at. \% C  \\
\hline\hline
5           &  2473                     &    $\beta$~B + graphite               &     5.617    & 12.160 & 332.251 & 12.8 - 16                   \\ \hline
5          &  2273                     &    $\beta$~B + amorph. C              &     5.634    & 12.125 & 333.41 & 12.8 - 16                  \\ \hline
5           &   1873 	         &    amorph. B + amorph. C	&     5.628   &    12.091 & 331.69 & 12.8 - 16  	         \\\hline
2           &  2273                     &    $\beta$~B + graphite               &     5.613    & 12.091 & 329.941 & 16 - 17.4                  \\ \hline
2           &	2073 	&	$\beta$~B + amorph. C		&	5.600 & 12.114 & 329.095 & 17.4 - 18.7	\\ \hline
2           &  1873                      &    amorph. B + amorph. C               &     5.596 & 12.079 & 327.628 & \textgreater 18.7   \\\hline
%This work   &    2 $\pm$ 0.2 GPa		 &   $\beta$~boron + graphite		  &        2273 $\pm$ 20 K			       \\ \hline
%This work   &    5 $\pm$ 0.2 GPa		 &   $\beta$~boron + amorphous carbon		  &        2273 $\pm$ 20 K	   \\\hline
%This work  &	5 $\pm$ 0.2 GPa	&	amorphous boron + amorphous carbon		&		2273 $\pm$ 20 K			\\ \hline
%This work  &	5 $\pm$ 0.2 GPa		&	 $\beta$~boron + graphite		&		2473 $\pm$ 20 K			\\\hline
\end{tabular}
%\captionsetup{width=17cm}
\caption{\label{tab:carbonconcentration} Atomic carbon concentration of the synthesised boron carbides as a function of the hexagonal cell parameters ($a$ and $c$) and the rhombohedral cell value ($V$) in ascending order of carbon concentration. The reference values for determining the carbon concentration from the cell volume were taken from Ref.~\cite{Aselage1992} and~\cite{Gosset2020}.}
%\end{landscape}
%\end{center}
\end{table*}
%\end{widetext}

\subsection{Effect of thermal cycling versus constant temperature increase}

During the \textit{in situ} experiment at 13 GPa, it was found that boron carbide formed in the last cycle when the temperature was raised to 2433~K when temperature was increased through thermal cycling. When the same experiment was repeated without the thermal cycling in \textit{ex situ} conditions at 2473~K, no boron carbide was formed. In both cases, the measured pressure was identical (within 0.5 GPa). This established that the synthesis temperature of boron carbide from elements at 13 GPa is above 2473~K, when steady heating is used. Hence, it is evident that the thermal cycling had played a role in the synthesis of boron carbide from its elements at a lower temperature. 

Some work has been done in the literature on the effects of thermal cycling in chemical reactions. Barratt \textit{et al.} \cite{Barratt2010} have shown through computational models that some hypothetical processes generate higher yields under forced thermal cycling than under single, fixed temperature conditions. Thermal cycling can also hasten phase transitions in some cases, as shown by Lee \textit{et al.} \cite{Lee2000}. They have shown the kinetic transition of $\gamma$ Fe-17 wt \% Mn alloy into $\epsilon$ phase due to thermal cycling: this can be attributed to the increase in transformation kinetics at the numerous nucleation sites introduced through thermal cycling.

Luk'yanenko \textit{et al.} \cite{Lukyanenko1997} have performed both theoretical and experimental investigations into the effect of thermal cycling on the high temperature interaction of titanium with gases like oxygen and nitrogen. They report that  thermal cycling had intensified the diffusion displacements of impurity atoms. The diffusion coefficients of these oxygen and nitrogen impurity atoms increase both in the matrix and in the newly-formed phase. This had strongly affected the conditions of formation and growth of the new phase because the number of possible nucleation centers on the phase boundary had permanently increased with time and the energy barrier of the phase transformation has decreased as well. 

%Thus, it can be hypothesised that the damage created in the $\beta$ boron structure due to the thermal cycling has permitted easier diffusion of the carbon atoms in the structure, 
Thus, we find that the dilation/densification cycles applied to $\beta$~boron structure favours the diffusion of carbon atoms, leading to nucleation sites of boron carbide. This has resulted in the formation of boron carbide at a lower temperature than that would be needed by a steady fixed temperature condition. %The changes that this process would create in the EDXRD pattern of the reacting materials is difficult to observe in this specific case, because of the recrystallisation of $\beta$ boron peaks under HPHT conditions.

\section{Conclusions}
\label{conc}

In the present work, we have synthesised boron carbide at  2~GPa, 5~GPa and 13~GPa. The optimum synthesis parameters of boron carbide at pressures up to 5~GPa from elemental boron and carbon have been studied for the first time, using the large volume Paris-Edinburgh press. It is seen that the mechanism of elemental formation of boron carbide is heavily dependent on pressure, which affects the temperature of formation of boron carbide. Amorphous carbon has proven to be a better reactant than graphite for the synthesis of boron carbide, since it lowers the temperature of formation, both at 2~GPa and at 5~GPa. Amorphous boron has also followed this trend, since it results in lower temperature of formation of boron carbide both at 2 GPa and 5 GPa, when compared with $\beta$~boron.

We have found that the carbon concentration of boron carbide depends on the pressure value, and propose pressure as a means to control carbon concentration in boron carbides.

The synthesis experiments at the higher pressure value of 13 GPa have been performed for the first time. We have shown that the synthesis temperature of boron carbide from $\beta$ boron and amorphous carbon is at least above 2473 K, with steady heating. The temperature of synthesis of boron carbide has been shown to decrease with thermal cycling that allows easier diffusion of carbon into the boron structure. Thus, thermal cycling can be investigated as a tool in bringing down the synthesis temperature in other chemical reactions as well.

\begin{acknowledgments}
Supports from the DGA (France) and from the program NEEDS-Matériaux (France) are gratefully acknowledged. The authors thank Benoit Baptiste, Keevin Béneut, Ludovic Delbes, Antoine Jay, Hicham Moutaabbid, and Silvia Pandolfi for useful discussions. The X-ray diffraction platform in IMPMC is also acknowledged. The PhD fellowship for A. Chakraborti has been provided by the Ecole Doctorale of Institut Polytechnique de Paris. We acknowledge SOLEIL for the provision of synchrotron radiation facilities.
\end{acknowledgments}
\bibliographystyle{apsrev4-2}
\bibliography{ref}

\end{document}